\documentclass[twocolumn,fleqn]{svjour2}    
\smartqed  
\usepackage{graphicx}
\usepackage{natbib}

\begin{document}
\newcommand{\SgrA}{Sgr\,A$^\star$\,}
\newcommand{\rg}{ \ensuremath{r_{\rm g}} }
\newcommand{\rms}{ \ensuremath{r_{\rm ms}} }
\newcommand{\mdot}{\ensuremath{\dot{M}_{\rm a}}}
\newcommand{\mdotB}{\ensuremath{\dot{M}_{\rm B}}}
\newcommand{\faca}{\ensuremath{\frac{3GM \dot{M}_{\rm a} x^{-3}}{8\pi r^3_g}}}
\newcommand{\facb}{\ensuremath{\frac{3GM \dot{M}_{\rm a}}{8\pi r^3}}}
\newcommand{\facc}{\ensuremath{\frac{3GM \dot{M}_{\rm a} }{2r_g}}}
\newcommand{\LEdd}{\ensuremath{L_{\rm Edd}}}
\newcommand{\Ld}{\ensuremath{L_{\rm d}}}
\newcommand{\ed}{\ensuremath{\epsilon_{\rm d}}}
\newcommand{\Pj}{\ensuremath{P_{\rm j}}}
\newcommand{\ej}{\ensuremath{\epsilon_{\rm j}}}
\newcommand{\Pa}{\ensuremath{P_{\rm a}}}
\newcommand{\ea}{\ensuremath{\epsilon_{\rm a}}}
\newcommand{\etot}{\ensuremath{\epsilon_{\rm tot}}}
\newcommand{\erphi}{\ensuremath{\epsilon_{r \phi}}}
\newcommand{\ephiz}{\ensuremath{\epsilon_{\phi z}}}
\newcommand{\ent}{\ensuremath{\epsilon^{\scriptscriptstyle{\rm NT}}}}
\newcommand{\enzt}{\ensuremath{\epsilon^{\rm nzt}}}
\newcommand{\fnt}{\ensuremath{f^{\scriptscriptstyle{\rm NT}}(x)}}
\newcommand{\frphi}{\ensuremath{f^{\rm nzt}_{r \phi}(x)}}
\newcommand{\fphiz}{\ensuremath{f^{\rm nzt}_{\phi z}(x)}}
\newcommand{\fntr}{\ensuremath{f^{\scriptscriptstyle{\rm NT}}(r)}}
\newcommand{\frphir}{\ensuremath{f^{\rm nzt}_{r \phi}(r)}}
\newcommand{\fphizr}{\ensuremath{f^{\nzt}_{\phi z}(r)}}
\newcommand{\xms}{\ensuremath{x_{\rm ms}}}
\newcommand{\yms}{\ensuremath{y_{\rm ms}}}
\newcommand{\nuobs}{\ensuremath{\nu_{\rm obs}}}
\newcommand{\sigmaT}{\sigma_{\scriptscriptstyle \rm T}}
\newcommand{\ltapprox}{\raisebox{-0.5ex}{$\,\stackrel{<}{\scriptstyle
\sim}\,$}}
\newcommand{\gtapprox}{\raisebox{-0.5ex}{$\,\stackrel{>}{\scriptstyle
\sim}\,$}}
\newcommand{\msolyr}{\ensuremath{\, M_\odot \, {\rm yr}^{-1}}}

\title{Jet-Driven Disk Accretion in Low Luminosity AGN?}

\titlerunning{Jet-Driven Accretion in LLAGN}

\author{E. J. D. Jolley         \and
        Z. Kuncic 
}


\institute{E. J. D. Jolley \at
              School of Physics, University of Sydney, Sydney, NSW, Australia, 2006. \\
              \email{E.Jolley@physics.usyd.edu.au} 
           \and
           Z. Kuncic \at
              School of Physics, University of Sydney, Sydney, NSW, Australia, 2006. \\
              \email{Z.Kuncic@physics.usyd.edu.au}
}

\date{Received: date / Accepted: date}

\maketitle

\begin{abstract}
We explore an accretion model for low luminosity AGN (LLAGN)
that attributes the low radiative output to a low mass accretion
rate, $\mdot$, rather than a low radiative efficiency. In this model,
electrons are assumed to drain energy from the ions as a result of
collisionless plasma microinstabilities. Consequently, the accreting gas collapses to 
form a geometrically thin disk at small radii and is able to cool before reaching the black hole. 
The accretion disk is not a standard disk, however, because the radial disk structure is 
modified by a magnetic torque which drives a jet and which is primarily responsible for 
angular momentum transport. We also include relativistic effects. 
We apply this model to the well known LLAGN M87 and calculate the combined disk-jet steady-state 
broadband spectrum. A comparison between predicted and observed spectra indicates that M87 may be a 
maximally spinning black hole accreting at a rate of $\sim 10^{-3} \, M_\odot \, {\rm yr}^{-1}$. 
This is about 6 orders of magnitude below the Eddington rate for the same radiative efficiency. 
Furthermore, the total jet power inferred by our model is in remarkably good agreement with the value independently deduced from observations of the M87 jet on kiloparsec scales.
\end{abstract}

\section{Introduction}
\label{intro}

We summarise an accretion model for LLAGN that has 
previously been applied to \SgrA \citep{Jolley07}. In this model, described in more detail below, 
initially collisionless gas accretes at a very low rate and collapses into a geometrically thin disk at 
small radii as a result of wave-particle resonances that facilitate efficient electron-ion coupling. 
Magnetic coupling between the disk and a jet is self-consistently modelled by a magnetic torque with a 
prescribed radial profile. 
Here, we apply this model to M87.

The giant elliptical galaxy M87 (NGC 4486) is a LLAGN situated at a 
distance of $d = (16 \pm 1.2)$ Mpc \citep{Tonry01} 
in the Virgo cluster. 
It harbours a central supermassive black hole (SMBH) of mass $M = (3.2 \pm 0.9) \times 10^9 M_\odot$, 
with a rapidly rotating disk of ionised gas, consisent with a
keplerian thin disk \citep{Macchetto97}, accompanying a prominent one-sided jet first detected by \citet{Curtis18}. 
The nucleus has a luminosity of $\approx 10^{42} \, {\rm ergs \, s}^{-1}$ \citep{Biretta91} which is at least two orders of 
magnitude below the luminosity expected for a standard thin accretion disk accreting at the Bondi rate $\dot{M}_{\rm B} = 0.1 \msolyr$, 
as determined from \textit{Chandra} X-ray observations \citep{Matteo03}. The observed luminosity from the nucleus is likely to be less than the total accretion power, however, 
because a significant proportion of the available accretion energy is used to power the large observed jet, with a total kinetic power estimated to be as large as $2 \times 10^{43} \, {\rm ergs \, s}^{-1}$ \citep{Reynolds96}. 

The inflowing plasma in LLAGN is collisionless because the dimensionless
mass accretion rate is so low ($\dot m \equiv L/\LEdd \ll 1$) that there insufficient time for electrons and ions to come into equipartition via two-body processes before reaching the event horizon
\citep*{SLE76,Ichimaru77,Rees82}. If other coupling processes are unable to equilibrate the electrons and ions within the inflow timescale and if the ions are preferentially heated by viscous dissipation of the gravitational binding energy, then the resulting accretion flow cannot radiate its internal energy before reaching the hole. This leads to a Radiatively Inefficient Accretion Flow (RIAF) \citep{NarYi94}.
RIAF models for LLAGN attribute the low luminosity to a low radiative efficiency, $\epsilon$, rather than a low mass accretion rate, $\mdot$. Recent variations on the basic RIAF model consider a reduced accretion rate close to the black hole due to convective motions \citep{Quataert00} or outflows \citep{BlandBeg99}. A reduced $\mdot$ at small radii appears to be necessary, at least for \SgrA, where polarization measurements imply $\mdot \ltapprox 4 \times 10^{-8} M_\odot \, {\rm yr}^{-1}$ \citep{JPM06}.

M87 has a large ($\approx 2 \, \rm kpc$ long) relativistic jet. Unlike winds, 
relativistic jets cannot carry away
large amounts of mass, and therefore cannot be responsible for reducing the accretion rate close to the black hole. 
Furthermore, the mechanism by which the
jet is coupled to the underlying accretion flow has yet to be explicitly modelled,
so the effect of the jet on the radial structure of the
underlying accretion flow is not known. Here, we consider the
possibility that the inflowing gas in M87 and other LLAGN is accreting at a very low rate and cools and settles into a compact,
geometrically thin disk at small radii. We suggest that the collisionless accreting plasma can form a cool disk because plasma microinstabilities act immediately to reduce any
large temperature gradient between the electrons and ions. Wave-particle resonances can then
efficiently couple the electrons and ions, as proposed by
\citet{BegChi88} (see also \citealt{BKL00,Quat98,Gruzinov98,QuatGruz99,Blackman99}). 
The presence of cold molecular gas inside the Bondi radius \citep{Tan07} may indicate the presence of a cool, thin accretion disk. 
This is inconsistent with RIAF models which form a geometrically thick, hot ion torus geometry \citep{Rees82}. 

In what follows, we wish to determine whether the low luminosity of M87 (and other
LLAGN) can be attributed to a low $\mdot$ rather than a low
radiative efficiency. We proceed by considering a geometrically thin, cool, single temperature relativistic accretion disk that is modified by magnetohydrodynamic (MHD) stresses. The 
disk is coupled to a relativistic jet via an MHD torque acting across the disk surface. This model is 
described in detail in \citet{Jolley07}. 
The radiative efficiency of our cool disk
is somewhat lower than that of the standard Shakura-Sunyaev disk \citep{b9} as a result of 
efficient extraction of accretion power by the jet. 
In Section \ref{sec2}, we examine the physical conditions needed for a
cool disk. 
In Section
\ref{sec3}, we present a summary of the relevant equations for 
the modified disk flux and the steady-state spectrum resulting
from a jet magnetically coupled to the underlying accretion flow. We
compare the spectrum predicted by our coupled disk-jet model with the observed 
spectrum for M87 in Section \ref{sec4}. A discussion of this work 
and some concluding remarks are given in Section \ref{Discussion}.

%

\section{Conditions for a Cool Disk} \label{sec2}

Here, we describe the properties of collisionless accretion
flows when the assumption of a two-temperature plasma ceases to remain valid. 
The detailed derivations are presented in \citet{Jolley07}.

If there exists a mechanism to transfer internal energy from the ions
to the electrons on an inflow time
\citep{BegChi88,BKL00,Quat98,Gruzinov98,QuatGruz99,Blackman99}, then
the accretion flow geometry will deflate from an ion-pressure-supported, two-temperature
torus predicted by RIAFs to a much thinner, quasi-thermal structure. For a
thermal pressure supported flow at the electron virial temperature
the 
height-to-radius ratio is $h/r = (m_{\rm e}/m_{\rm p})^{1/2}
\approx 0.02$. This defines a geometrically thin disk. 
The condition that the ratio of the cooling timescale to the inflow timescale 
$t^{\rm cool} / t^{\rm inflow} \ltapprox 1$ is required for the disk to cool 
before the gas reaches the black hole. This implies \citep[see also e.g.][]{Rees82}
\begin{equation}\label{condition}
\dot m \gtapprox 2 \times 10^{-12} \epsilon_{0.1} \alpha_{0.1}^2 \left( \frac{r}{100\rg} \right)^{-1/2} \left( \frac{h}{10^{-2}r} \right)^6 \qquad ,
\end{equation}
where $\epsilon = 0.1 \epsilon_{0.1}$ is the radiative efficiency and $\alpha = 0.1 \alpha_{0.1}$ is the dimensionless viscosity parameter from standard accretion disk theory \citep{b9}.
Hence by relaxing the assumption that
electrons and ions can interact only through Coulomb collisions, the collisionless accretion flow in low-$\dot m$ systems must be geometrically thin. The condition (\ref{condition}) implies that the collisionless accretion flow is able to cool on an inflow timescale and thus, is \emph{not} advective.

\section{Coupling a Magnetized Jet to a Relativistic, Cool Disk}\label{sec3}

\subsection{Non-Zero Torque on the Disk Surface}\label{NZTatsurface}

The MHD stresses present in an accretion disk can produce a torque acting across the 
disk surface as well as at the last marginally stable orbit. The surface torque can efficiently remove angular momentum vertically outwards from the disk and does work against the disk, directing energy vertically to form a magnetized jet \citep{KB04}. 

The radiative flux from a relativistic, torqued disk is \citep{Jolley07}
\begin{equation}
F (r) = \frac{3GM\mdot}{8\pi r^3} \, \left[ \, \fnt + \frphi - \fphiz \, \right]
\label{fluxeq}
\end{equation}
where $x = r/\rg$, $\fnt$ is the \citet{b10} relativistic correction
factor, $\frphi$ is a correction factor for a nonzero torque (NZT) at
the inner disk boundary \citep{b12}, and $\fphiz$ is an analogous
correction factor for a nonzero torque on the disk surface. 

Using global energy conservation the disk flux profile (\ref{fluxeq}) 
can be expressed (see \citealt{Jolley07} for details):
\begin{eqnarray}
 F (r) = \frac{3GM\mdot}{8\pi r^3}
\left[ \frac{A(x)}{C(x)} + \frac{2}{3} \frac{\Delta \epsilon}{C(x) x^{1/2} I_3} \right. \nonumber \\
\left. -  \frac{2}{3} \frac{\ej}{C(x) x^{1/2}} \frac{I_1(x)}{I_2}  \right] 
\label{Fd}
\end{eqnarray}
where 
\begin{equation}
\Delta \epsilon = \frac{3}{2} \int_{\xms}^\infty  x^{-2} \frphi \, \mathrm{d}x
\end{equation} 
is the efficiency of the torque acting at the last marginally stable orbit, 
and 
\begin{equation}
\ej = \frac{3}{2} \int_{\xms}^\infty x^{-2} \fphiz \, \mathrm{d}x
\end{equation} 
is the efficiency of the jet. The functions $C(x)$ and $A(x)$ are relativistic correction factors 
from \citet{b10}, and the following are derived in \citet{Jolley07}:
\begin{equation}
I_1(x) = \int_{\xms}^x \frac{[C(x)]^{1/2}}{B(x)} x^{2-q} \, {\rm d}x
\end{equation}
\begin{equation}
I_2 = \int_{\xms}^{\infty} \frac{x^{-5/2}}{C(x)} \, I_1(x) \, {\rm d}x
\end{equation}
\begin{equation}
I_3 = \int_{\xms}^\infty \frac{x^{-5/2}}{C(x)} \, {\rm d}x
\end{equation}
The mass accretion rate can be written as
\begin{equation}
\mdot = \left( \frac{\Ld}{\LEdd} \right) \frac{\LEdd}{\ed c^2}
\end{equation}
where $\LEdd = 4\pi GM m_{\rm p} c / \sigmaT$. 

The input parameters for the modified disk model are the dimensionless black hole spin parameter $a$, 
the fractional efficiency of the torque at the last marginally stable orbit 
$\Delta \epsilon/\ent = 0.10$, 
and the fraction of accretion power injected into the jet, 
$\ej / \ea$. This last parameter has an upper limit in order 
for the disk flux to remain positive at all radii.

\begin{figure}
\centerline{\includegraphics[height = 12truecm]{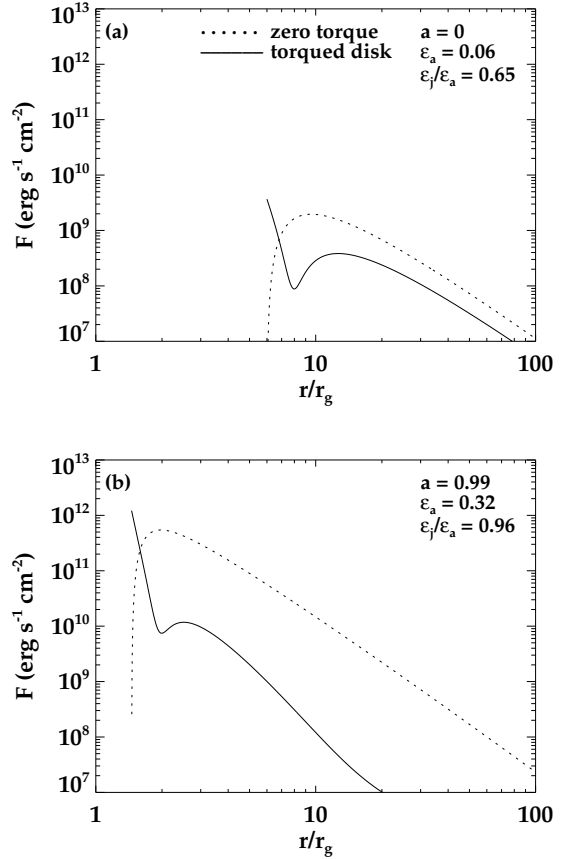}}
\caption{Radial flux profiles for the jet-modified disk model with different model parameters: 
$a$ is the black hole spin parameter, $\ea$ is the overall accretion efficiency 
and $\ej/\ea$ is the fractional jet power. 
The solid line corresponds to a relativistic disk torqued at the inner boundary and on its surface; the dotted line corresponds to the same disk without torque effects.}
\label{fluxfig}
\end{figure}

The effect of the nonzero torque
across the disk surface is to do work against the disk, thus reducing
the disk flux over the range of radii where the magnetic torque is
strongest. This counteracts the effect of the nonzero torque at the
inner disk boundary, which enchances the disk flux near $\xms$.
The
combined effects of these two torques is clearly evident in the radial
flux profiles in Fig.~\ref{fluxfig}. The resulting disk flux
radial profiles are substantially modified from their corresponding
zero-torque profiles (Fig.~\ref{fluxfig}, dotted lines). 
It is clear that 
the nonzero magnetic torque 
acting on the disk surface results in a disk radiative efficiency $\ed$ 
that is lower than that of a non-torqued disk. 

We have explicitly taken into account how 
the \textit{local} disk radial structure is modified by a magnetized jet that is primarily responsible for angular momentum 
transport. This 
results in a disk spectrum that is modified with respect to that predicted by standard theory. 

%
\subsection{Jet Emission}

We identify the nonzero magnetic
torque across the disk surface with the mechanism responsible for
extracting accretion power from the disk and injecting it into a
jet. 
Some of the magnetic energy is subsequently converted into kinetic energy. 
We expect a fraction of the particles to be accelerated to nonthermal, relativistic energies. 
Synchrotron radiation by relativistic electrons will then contribute significantly to the 
radio emission.

Following the method in \citet{Jolley07}, we divide the jet into a series of quasi-cylindrical sections of thickness $\Delta z$, 
and calculate the total emission spectrum 
by summing the contributions from each component.


We consider a relativistic jet with bulk Lorentz factor $\Gamma_{\rm j}$ and Doppler factor 
$\delta = \left\{ \Gamma_{\rm j} \left[1 - (1-\Gamma_{\rm j}^{-2})^{1/2} \cos \theta_{\rm j} \right] \right\}^{-1}$,
where $\theta_{\rm j}$ is the angle between our line of sight and the M87 jet axis. 
We use the following simple radiative transfer model to calculate the observed specific 
luminosity due to the net contribution from each jet component 
(assuming isotropic emission in the source rest frame): 
\begin{equation} \label{L}
L_{\rm \nuobs}^{\rm obs} \approx 2\sum^{z_{\rm j}}_{z = z_0}  
4 \pi \delta^3 S_{\rm \nuobs}^{\rm syn}\left( 1- {\rm e}^{-\tau_{\rm \nuobs}^{\rm syn}} \right) \Delta A 
\end{equation}
where $\Delta A \approx \pi r \Delta z \sin \theta_{\rm j}$ is the projected surface area of each emitting cylinder, 
$S_{\rm \nuobs}^{\rm syn}$ is the synchrotron source function (see e.g. \citealt{RL} for relevant formulas) and 
$$\tau_{\rm \nuobs}^{\rm syn} = \delta^{-1} \kappa_{\rm \nuobs}^{\rm syn} \Delta s$$ 
is the synchrotron optical depth along a path length 
$\Delta s$ through each cylindrical section. 

The electron number density $N_{\rm e}$ and hence the magnetic field $B$ 
decline with jet height $z$ according to 
$$N_{\rm e}(z) \propto z^{-2} \qquad {\rm ,} \qquad B(z) \propto z^{-1}$$
The total jet power is 
\begin{eqnarray}
\Pj \approx \pi r_{\rm j}^2 \Gamma_{\rm j} (1 - \Gamma_{\rm j}^{-2})^{1/2} {\rm c} \left[ (\Gamma_{\rm j} - 1)N_{\rm e} m_p {\rm c}^2 \right.  \nonumber \\
\left. + \frac{4}{3} \Gamma_{\rm j} N_{\rm e} \langle\gamma\rangle m_{\rm e} {\rm c}^2 \left( 1+ 2f_{\rm eq} \right) \right]
\end{eqnarray}
where the first term in square brackets refers to the bulk jet kinetic energy 
and the second term refers to the electron kinetic energy and the magnetic energy. The equipartition factor $f_{\rm eq} = 1$ is used 
to relate the magnetic and electron energy densities. Equation (11) is used to calculate $N_{\rm e}$ at the base of the jet. 

\section{Application to M87}\label{sec4}

\begin{figure}
\begin{center}
  \includegraphics[height = 12truecm]{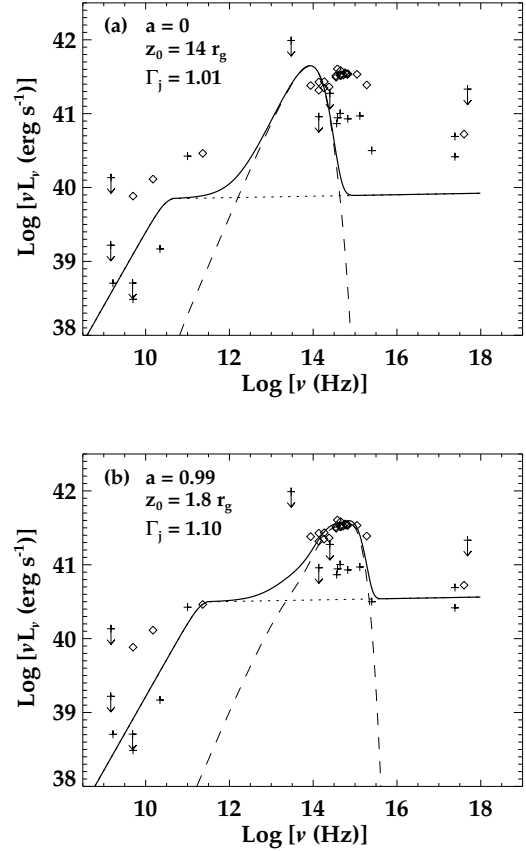}
\end{center}
\caption{Observed and predicted spectra for M87. The plus symbols are data points from
\citet{Ho99} and the diamonds are from Meisenheimer et. al. (2007). 
The solid line is the theoretical disk+jet steady-state spectrum predicted by our model. 
The dashed line is the disk spectrum, and the dotted line is the synchrotron jet spectrum. 
(a) is for a  
black hole with zero spin, and (b) is for a maximally spinning black hole. The launching height 
and the bulk Lorentz factor for the jet are given by $z_0$ and $\Gamma_{\rm j}$, respectively.
See Table 1 for other parameters.}
\label{datafig}       
\end{figure}

Figure~\ref{datafig} shows the predicted combined disk and jet spectra for our model for different parameters.
The observational data points (corrected for extinction) are taken from \citet{Ho99} (plus signs) and \citet{Meis96} (diamonds). 
The inclination of the nuclear disk rotation axis to our line of sight is $\theta_{\rm d} \approx 60^\circ$ \citep{Macchetto97}, 
and the jet inclination angle is $\theta_{\rm j} \approx 30^\circ$, with an opening angle of $\approx 60^\circ$ \citep{Ly07} 
(see e.g. \citealt{Biretta99} 
for a review of the properties of the M87 jet). The disk luminosity as a fraction of the 
Eddington luminosity is 
$L_{\rm d} / L_{\rm Edd} = \dot{m} = 1.8 \times 10^{-6}$.
Table~1 lists the other physical parameters used in our model for the zero spin ($a = 0$) and 
maximally spinning ($a \approx 1$) black hole cases. 


\begin{table}
\label{table1}       
\caption{Parameters used in the disk model. The black hole spin is $a$,
$\ej/\ea$ is the maximum allowable fraction of the accretion power removed by the jet,  
$\ea$ is the total accretion efficiency, 
$\ed$ is the disk radiative efficiency, $\mdot$ is the accretion rate in $M_{\odot}{\rm yr}^{-1}$ and 
$P_{\rm j}$ is the total jet power in erg s$^{-1}$. 
The black hole mass is $M = 3.2 \times 10^9 M_\odot$, 
the disk luminosity as a fraction of the Eddington luminosity is $\dot{m} = 1.8 \times 10^{-6}$ and 
the optically-thin synchrotron spectral index is $\alpha = 0.99$.}
\begin{tabular}{cc|cccc}\hline\hline
\multicolumn{2}{c}{input parameters} & \multicolumn{4}{c}{inferred parameters}\\
\hline\noalign{\smallskip}
a&$\ej/\ea$&$\ea$&$\ed$&$\mdot$&$P_{\rm j}$ \\
\noalign{\smallskip}\hline\noalign{\smallskip}
$0.00$&$0.65$&$0.06$&$0.02$&$6 \times 10^{-2}$&$1 \times 10^{42}$\\
$0.99$&$0.96$&$0.32$&$0.01$&$1 \times 10^{-3}$&$2 \times 10^{43}$\\
\noalign{\smallskip}\hline
\end{tabular}
\end{table}
 
An accretion disk around a spinning black hole can extract more accretion power, $P_{\rm a} = \ea \mdot c^2$,
than a disk around a non-spinning hole for the same accretion rate. 
This is because the last marginally stable orbit for $a = 0.99$ is 
much smaller than that for $a = 0$ (where $r_{\rm ms} \approx 6 r_{\rm g}$). 
Thus, the accretion disk in the high-spin case (with $r_{\rm ms} \approx 1.45 r_{\rm g}$ for $a = 0.99$) 
reaches higher temperatures and emits a bluer spectrum than that in the zero spin case. 

%

Similarly, the jet launching height $z_0$ determines the normalization of the relativistic electron 
number density $N_{\rm e}$ and hence the jet synchrotron spectrum. Both $z_0$ and $\Gamma_{\rm j}$ are 
constrained by the requirement that the jet remains non-dissipative (i.e. radiatively inefficient) and 
relativistic. These jet constraints produce a better overall agreement with the data for the high-spin case. 
Furthermore, the jet power predicted for the maximally spinning ($a = 0.99$) case is in excellent agreement with the value 
$P_{\rm j} \approx 2 \times 10^{43} \, {\rm erg s}^{-1}$ deduced by \citet{Reynolds96} from observations 
of the M87 jet on kiloparsec scales.
Our model thus predicts that the black hole in M87 may be maximally spinning.

%


\section{Discussion and Conclusion}\label{Discussion}

We have presented a model which combines existing theory for relativistic disk accretion with an
explicit prescription for disk-jet coupling via a magnetic torque on
the disk surface. The torque efficiently extracts angular momentum and energy from the disk at
small radii to drive a magnetized jet. 
Using this model, we have demonstrated that the low radiative output from M87 can be attributed to a low mass accretion 
rate rather than a low radiative efficiency. 
From the predicted combined jet and disk spectra, 
our model indicates that M87 may be a rapidly spinning black hole with a dimensionless spin $a \approx 0.99$. 
We predict a mass accretion rate $\mdot \approx 1 \times 10^{-3} \,{\rm M_{\odot}\, yr^{-1}}$ 
and a disk radiative efficiency $\ed \approx 0.01$.  
This interpretation of the nature of black hole accretion in M87 differs from 
that of radiatively inefficient models, which attribute the low luminosity to an unusually low radiative 
efficiency, typically $\ltapprox 10^{-5}$. 
Whereas radiatively inefficent accretion flows assume a thick, bloated torus geometry with a hot two-temperature plasma, 
our model assumes a geometrically thin, quasi-thermal disk. 

Our model for low-$\dot m$ accretors requires an efficient thermal 
coupling mechanism between the electrons and ions facilitated by collisionless plasma instabilities, 
resulting in a
geometrically thin disk which can cool on an inflow timescale. The exact 
nature of the collisionless plasma instabilities required is currently 
the subject of future work.

There is observational evidence for the presence of cold gas 
in the vicinity of the nuclear disk in M87, in the form of molecular 
gas inside the Bondi radius. There has been 
speculation that the mass accretion rate could be reduced due to star formation in these regions, 
although any definitive 
evidence for this scenario has yet to be found \citep{Tan07}.

%
%
%
\begin{acknowledgements}
E. J. D. Jolley acknowledges support from a University of Sydney Postgraduate Award. \\
Z. Kuncic acknowledges support from a University of Sydney Research Grant. 
\end{acknowledgements}

\bibliographystyle{natbib}



\end{document}